\documentclass[two column, prl]{revtex4}
\usepackage{graphics,graphicx}
\usepackage{ulem}
\begin{document}
\title{Magnetic order and instability in newly synthesized CoSeAs marcasite}
\author{Y. Chen$^{1}$}
\author{G. Yumnam$^{1}$}
\author{A. Dahal$^{1}$}
\author{J. Rodriguez$^{2,3}$}
\author{G. Xu$^{2}$}
\author{T. Heitmann$^{4}$}
\author{D. K.~ Singh$^{1,*}$}
\affiliation{$^{1}$Department of Physics and Astronomy, University of Missouri, Columbia, MO 65211}
\affiliation{$^{2}$National Institute of Standards and Technology, Gaithersburg, MD 20899}
\affiliation{$^{3}$Department of Materials Sciences and Engineering, University of Maryland, College Park, MD 20742}
\affiliation{$^{4}$University of Missouri Research Reactor, Columbia, MO 65211}
\affiliation{$^{*}$email: singhdk@missouri.edu}

\begin{abstract}
Marcasite class of compounds provide facile platform to explore novel phenomena of fundamental and technological importance, such as unconventional superconductivity or high performance electrocatalyst. We report the synthesis and experimental investigation of a new marcasite CoSeAs in this letter. Experimental investigation of the new material using neutron scattering measurements reveal weak magnetic correlation of cobalt ions below $T$ = 36.2 K. The modest isotropic exchange interaction between cobalt moments, inferred from random phase approximation analysis, hints of magnetically unstable environment. It is a desirable characteristic to induce unconventional superconductivity via chemical pressure application. 
\end{abstract}

\maketitle

Transition metals are at the core of almost every novel phenomena in magnetism.\cite{Zaanen,Carlin} Among the many chemical groups of transition metal compounds, marcasite phase is one of the most intriguing lattice groups with FeS$_{2}$-type crystal structure.\cite{marcasite,marcasite2} These materials are of strong technological importance, especially for application in photovoltaics and in the design of efficient electrocatalyst. Some of the notable examples include the demonstration of high performance electrocatalyst in FeX$_{2}$ ($X$ = S, Se) and high absorption coefficient photovoltaic property in CoSe$_{2}$.\cite{Chen,Zhang,Li} Marcasites are also known to manifest superconducting phenomena of both conventional and unconventional origins.\cite{Hull,Amsler} The proposed observation of unconventional superconductivity in marcasite phase FeBi$_{2}$ is attributed to the competing instability of underlying magnetism in the system.\cite{Amsler} The chemical structure of a marcasite is described by the distorted octahedrons of six anions (ligands) enveloping the cation of 3d or 4d transition metal ion.\cite{Endo} Despite the presence of transition metal as the key constituting element in the stoichiometric composition, most of them are either diamagnetic or paramagnetic with semiconducting electrical characteristic.\cite{Hulliger} The Jahn-Teller distortion in transition metal ion coordination octahedron is arguably responsible for the non-magnetic ground state in majority of the transition metal marcasites.\cite{Hull} We have synthesized a new compound CoSeAs in this series. Detailed experimental investigations of CoSeAs using elastic and inelastic neutron scattering measurements suggest the development of long range magnetic order below $T_c$ = 36.2 K. It sets a new precedent in 1:1:1 stoichiometric composition of the corresponding lattice group. Furthermore, we find that Co ions are correlated with the weak nearest neighbor exchange interaction, $J$ = 0.25 (4) meV, which makes the system susceptible to a transition to non-magnetic or different phase of matter, such as superconductivity, under modest external effect.

CoSeAs stands at the cross-roads of CoSe and CoAs compounds that crystallize in MnP-type tetragonal structure. While CoSe is argued to manifest a combination of ferromagnetic and spin glass properties,\cite{Efrain,Paglione} CoAs is considered non-magnetic.\cite{Selte} However, both materials exhibit the metallic characteristic. Sharing magnetic traits of both compounds, CoSeAs is on the verge of magnetic instability. CoSeAs crystallizes in the FeS$_{2}$-type marcasite structure with weak metallic characteristic, bordering to the semiconducting phase at low temperature. We have synthesized the polycrystalline samples of CoSeAs using repetitive solid-state reaction method in evacuated quartz tubes. The starting materials were 99.998\% Co, 99.999\% Se (Alfa Aesar) and 99.997\% As (Sigma-Aldrich).\cite{NIST} CoSe was synthesized from the stoichiometric composition of Co and Se. The mixture was grinded, pelletized and loaded into a quartz tube, subsequently evacuated and sealed, then annealed at 900$^{o}$ C for two days. After confirming the pure structure of CoSe (with no oxidation) using X-ray diffraction (XRD) measurement, stoichiometric amount of As was added to CoSe powder. The mixture was grinded, pelletized and sintered in evacuated quartz tube at 900$^{o}$ C for another two days. As shown in Fig. 1, the XRD pattern clearly manifests the high quality of as synthesized polycrystalline CoSeAs. X-ray peaks are well indexed by the $Pnnm$ space group body centered orthorhombic FeS$_{2}$-type marcasite structure, as shown in the inset in Fig. 1, with lattice parameters of $\textit{a}$ = 4.756 $\AA$, $\textit{b}$ =5.756 $\AA$ and $\textit{c}$ =3.570 $\AA$.

\begin{figure}
\centering
\includegraphics[width=8.8 cm]{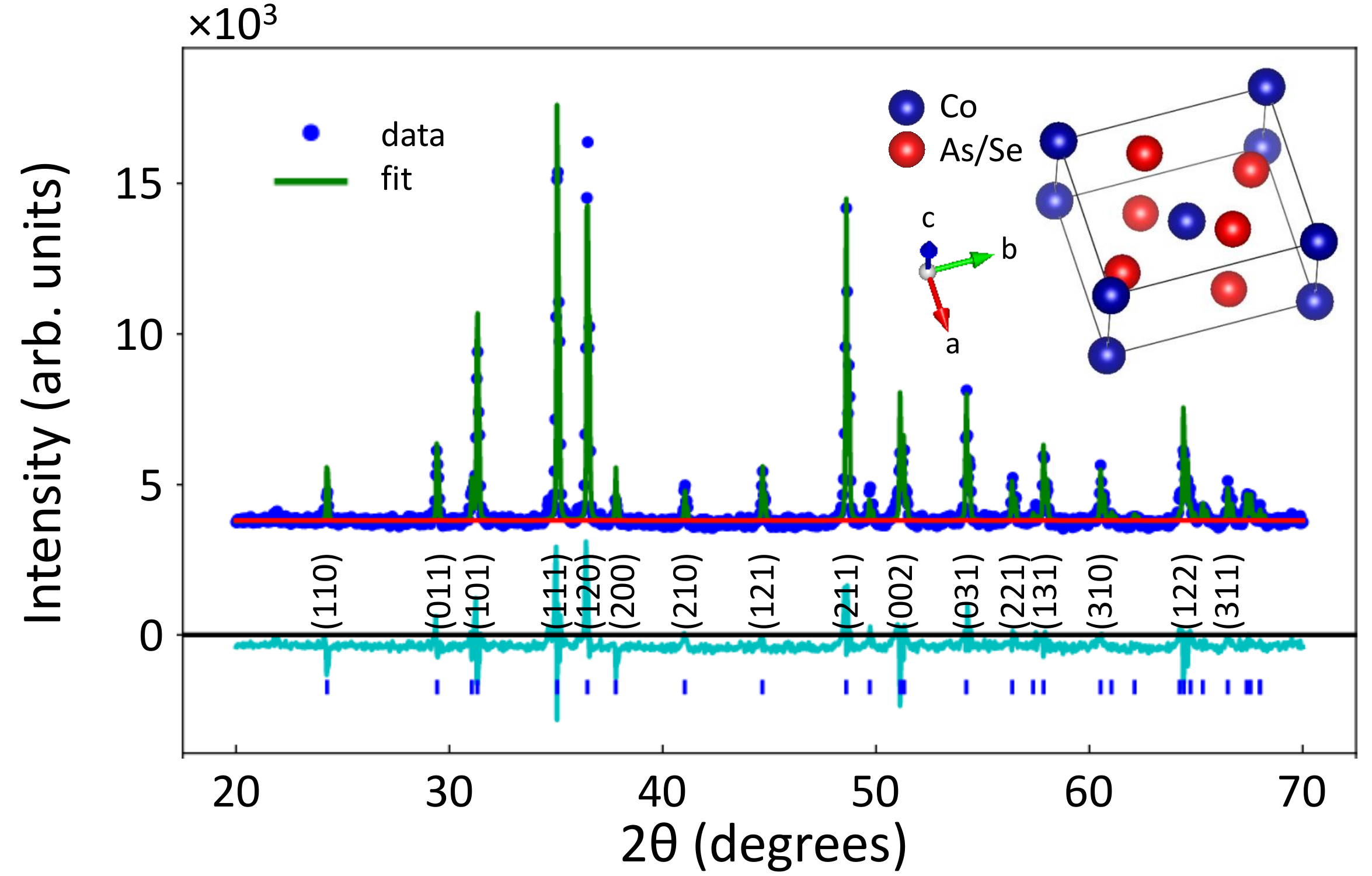} \vspace{-6mm}
\caption{(color online) Crystal structure of CoSeAs marcasite and X-ray diffraction data. X-ray diffraction pattern of as synthesized CoSeAs. XRD peaks are indexed with the FeS$_{2}$-type marcasite structure. Inset shows the crystal structure of CoSeAs compound.
} \vspace{-6mm}
\end{figure}

There is not much known about CoSeAs. The knowledge of electrical conducting properties and a theoretical understanding of the density of states at the Fermi surface are necessary to characterize the new material. We have performed first principles electronic structure calculations for CoSeAs based on the density functional theory by employing the plane-wave basis set, as implemented in the Quantum-ESPRESSO.\cite{Baroni} The projector augmented wave method was used with Troullier-Martins norm-conserving pseudopotential with nonlinear core correction. The calculations were performed with (without) spin-orbit interaction by using fully (scalar) relativistic pseudopotentials. The exchange correlation functional was treated within the generalized gradient approximation of Perdew-Burke-Ernzerhof (PBE-GGA).\cite{Perdew} The correlation effects of Co 3$d$ electrons were included via GGA+U method within the simplified rotational invariant scheme of Cococcioni et al.\cite{Coco} The value of on-site Coulombic interaction term (U) was set to a well-tested value of $U$ = 4.5 eV. A well converged kinetic-energy cutoff of 80 Ry was used with a Monkhorst-Pack sampling of 16$\times$16$\times$16. The experimental lattice parameter obtained from X-ray diffraction measurements were used as the initial configuration of the atoms for the DFT calculations. The magnetic configuration of the Co atoms was initialized to be ferromagnetic. Note that the final magnetic configuration of Co atoms is independent of the initial magnetization direction. A strict self-consistent energy convergence criterion of 10$^{-8}$ Ry was imposed. As shown in Fig. 2c, the calculated spin-resolved density of states (DOS) show that the characteristic Co-$d$ states are embedded well below the fermi energy for both majority (Fig. 2a) and minority (Fig. 2b) spin carriers, indicating weak-metallic or semiconducting characteristic due to the embedded Co d-orbitals. We note that the minority-spin carriers exhibit a much larger DOS at the Fermi level than the majority-spin carriers, which is a typical characteristic of ferromagnetic material. This is further substantiated by the spin-resolved hole Fermi surface, as shown in Fig. 2b, where the minority-carrier hole Fermi surface is more populated even though it has a much smaller volume compared to that of the majority-carriers (Fig. 2a). It gives rise to non-degenerate energies of electrons with opposite spins.

The high quality polycrystalline sample of CoSeAs was electrically characterized using a closed cycle refrigerator based cryostat with a base temperature of $T \simeq$ 5 K. As shown in Fig. 2d, the system exhibits very weak metallic-to-semiconducting behavior as temperature is reduced, also consistent with the DFT calculations. Information about the underlying magnetic properties is obtained from detailed elastic and inelastic neutron scattering measurements. Neutron measurements were performed on a $~$4.4g polycrystalline sample of CoSeAs at the multi-axis crystal spectrometer (MACS) with fixed final neutron energy of $E_f$ = 3.7 meV at the NIST Center for Neutron Research. Additional measurements were performed on the spin-polarized triple-axis spectrometer (SPINS) at NCNR and a position sensitive detector (PSD) powder diffractometer at the University of Missouri Research Reactor with fixed final neutron energy of 37 meV using graphite monochromator. Elastic measurements at SPINS employed a flat pyrolytic graphite (PG) analyzer with cold BeO filter in front, while the measurements on the powder diffractometer was performed using the tighter collimations before the monochromator and a PG filter. Inelastic measurements on MACS were performed in the focussed analyzer configuration with fixed $E_F$ = 3.7 meV and the energy resolution of $\simeq$ 0.25 meV. The sample was loaded in liquid $^{4}$He cooled cryostat with the lowest accessible temperature of $T$ = 1.7 K. 

\begin{figure}
\centering
\includegraphics[width=8.7 cm]{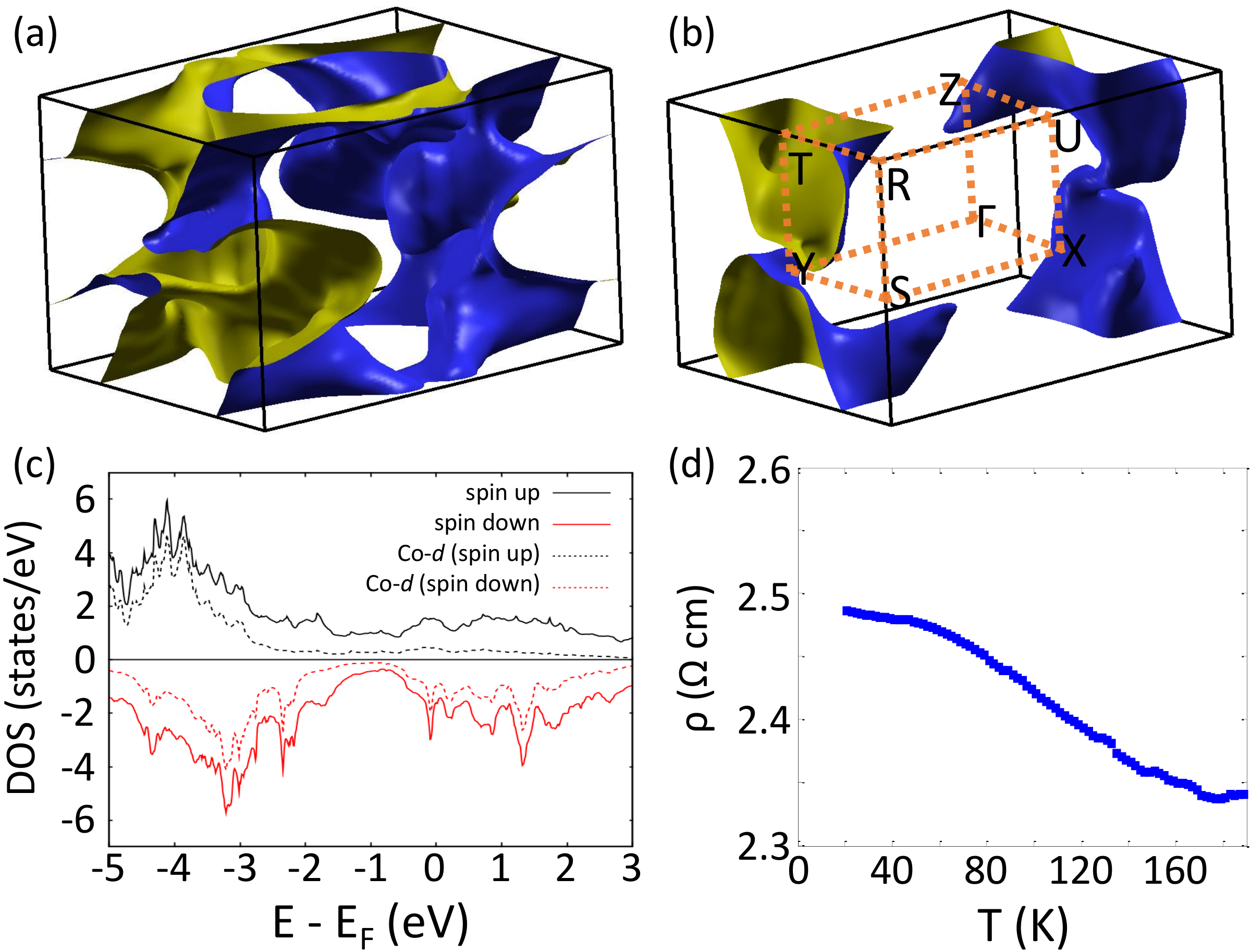} \vspace{-6mm}
\caption{(color online) Fermi surface and electrical characteristic of CoSeAs. (a-b) Spin resolved Fermi surfaces of CoSeAs: spin up (Fig. a) and spin down (Fig. b). (c) Density of states calculated using density functional theory (see text) elucidates the semiconducting character of the material. (d) Electrical measurement shows very weak metallic conductivity at high temperature. The system manifests semiconducting behavior at lower temperature, which is in agreement with electrical properties of marcasite phase compounds.
} \vspace{-6mm}
\end{figure}

\begin{figure*}
\centering
\includegraphics[width=18 cm]{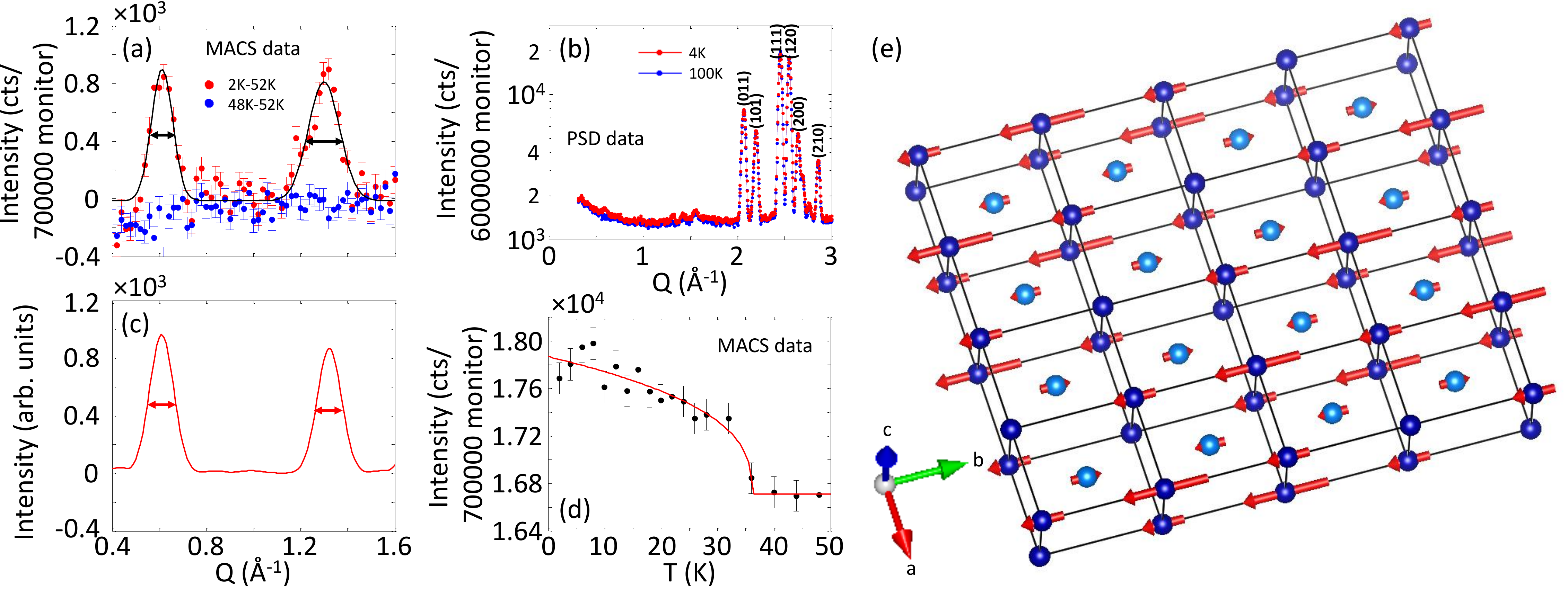} \vspace{-4mm}
\caption{(color online) Elastic neutron scattering measurements of CoSeAs and order parameter as a function of temperature. (a) Two-theta scans (converted into absolute wave vector q) at different temperatures, obtained on MACS spectrometer. Experimental results were also reproduced on SPINS spectrometer. Long monitor count was implemented to resolve the weak underlying magnetism in the system. Magnetic Bragg peaks, identified as (1/4 1/4 1/4) (equivalent to q = 0.6 $\AA$$^{-1}$) and (100) (equivalent to q = 1.32 $\AA$$^{-1}$) in the figure, emerge at low q as temperature reduces. Experimental data is well described by resolution limited Gaussian curve. (b) Two-theta scans spanning a broader q range, performed at PSD instrument at MURR. Magnetic peaks are about ten times weaker than the weakest nuclear peak. (c) Numerically simulated peak intensity for the spin structure, shown in Fig. e, is used to construct the Gaussian profile by utilizing the instrument's q resolution. Calculated value is in good agreement with experimental results. (d) Intensity vs. temperature at magnetic q = 0.595 $\AA$$^{-1}$. Magnetic order transition temperature of $T$ = 36.2 K is deduced by fitting the order parameter plot using power law. (e) Spin configuration of Co ions in CoSeAs. In all plots, error bars represent one standard deviation. } \vspace{-6mm}
\end{figure*}

Elastic scattering measurements on CoSeAs powder are used to infer the underlying static magnetic correlation between Co ions. We show the representative scans at two temperatures in Fig. 3a. Additional Bragg peaks arise as the sample is cooled to low temperature, indicating the development of magnetic order in the system. Compared to the nuclear peak intensities, shown in Fig. 3b, magnetic peaks are significantly weaker. Furthermore, elastic measurements required long counting time to obtain the statistically significant magnetic peak intensities. Together, they hint of small ordered moment of Co ions in the system.\cite{Harriger} Experimental data is well described by the resolution limited Gaussian lineshape. The estimated full width half maximum (FWHM) of magnetic peak is comparable to the instrument resolution of MACS spectrometer, suggesting the existence of long range magnetic order in CoSeAs. The magnetic order is found to persist to reasonably high temperature. In Fig. 3d, we show the plot of order parameter as a function of temperature at the magnetic wave vector q = 0.595 $\AA$$^{-1}$. Fitting of experimental data using a power law, given by $I$ $\propto$ (1-T/T$_{c}$)$^{-\beta}$,\cite{Singh} yields a transition temperature of $T_c$ = 36.2 K to magnetic ordered state. The estimated value of power law exponent is $\beta$ = 0.357(4). 

The magnetic wave vectors are identified to be both integer and rational fractions of reciprocal lattice units e.g. (100) and (1/4 1/4 1/4). The nature of magnetic correlation is deduced by performing detailed numerical modeling of experimental data. The experimentally observed structure factor, estimated from the Gaussian fit of the elastic data, are compared with the numerically calculated structure factor for model spin configurations. Structure factor is calculated using, $ F_{M} =\sum_{j} S_{\perp j} p_{j}  e^{ iQr_{j} }e^{-W_{j}} $,\cite{Shirane} where $ S_{\perp} =\hat{Q}\times(S\times\hat{Q})$ is the spin component perpendicular to the Q, $ p = (\frac{\gamma r_{0}}{2})gf(Q) $, $ (\frac{\gamma r_{0}}{2}) $= 0.2695 $ \times 10^{-12} cm$, g is the Lande splitting factor and was taken to be g = 2, $ f(Q) $ is the magnetic form factor, and $e^{-W_{j}} $ is the Debye-Waller factor and was taken to be 1.\cite{Shirane,Dahal2} Simulated intensities are powder averaged by multiplying with an appropriate factor of (1/sin($\theta$).cos(2$\theta$)).\cite{Shirane} Best fit to experimental data is obtained for magnetic moments arrangement comprised of two magnetic sublattices: (a) Co ions occupying the vertices of the orthorhombic lattice are ferromagnetically aligned along the $b$-axis and arranged in a density wave configuration with quadrupled magnetic unit cell, and (b) Co moment at the body-centered position are arranged in density wave configuration with quadrupled magnetic unit cell (see details in Supplementary Materials). As shown in Fig. 3c, the numerically simulated powder profile for the aforementioned spin structure, shown in Fig. 3e, well describes the experimentally observed diffraction pattern. The proposed spin configuration also gives rise to peak intensity at (001) rlu, which was not resolved in the experimental data. The estimated ordered moment is 0.26(5)$\mu_B$. Such a small value of the ordered moment reflects the nearly compensated spin polarities in individual anion octahedron. This is consistent with the general observation in marcasite phase magnetic material of weak or no magnetic order due to the strong screening of magnetic moment by conduction electrons. Experimental findings hint of the weakly correlated Co ions in CoSeAs. 

To gain insight about the strength of exchange interaction between Co ions in CoSeAs, we have performed detailed inelastic measurements. In Fig. 4a, we show the color map of inelastic spectrum, obtained on MACS spectrometer, at $T$ = 1.7 K. A q-independent band of inelastic excitation tends to develop below E $\simeq$ 3 meV at low temperature. The excitation at higher q follows the Co form factor, thus gradually weakens. Inelastic data is background subtracted and thermally balanced by multiplying the Intensity by a factor of $\pi$(1-exp(-$E$/$k_B$$T$)). The absence of any dispersion in experimental data suggests an isotropic nearest neighbor interaction in the system. Further quantitative information is obtained by analyzing the dynamic susceptibility $\chi$$^{''}$(Q, E), given by

\begin{eqnarray}
{S(Q, \omega)}&=&{\gamma_0}^{2}(\frac{k_i}{k_f}){f(Q)}^{2}\frac{1}{1- {e}^{-h\omega/{k_B}T}}(\frac{{\chi}^{^{\prime\prime}}(Q, \omega)}{\pi})
\end{eqnarray}

where $\gamma_0$$^{2}$ = 0.073$/$$\mu_B$$^{2}$, $k$$_{i}$ and $k$$_{f}$ represent initial and final neutron wave vectors and $f(Q)$ is the form factor of magnetic ion (in this case Co ion). In Fig. 4c, we plot $\chi$$^{''}$(Q, E) as a function of energy at $T$ = 1.7 K at a representative Q = 1.4 $\AA$$^{-1}$. Clearly, a broad peak in $\chi$$^{''}$(Q, E), centered around $E$ $\simeq$ 0.75 meV, is observed. At higher temperature, above $T_c$, the excitation becomes indistinguishable from the background, indicating magnetic nature of inelastic peak. A dispersive excitation is detected at higher temperature, which can be associated to the phonon excitation in the system. However, the excitation does not prevail to high q values, typical of phonon excitation. One explanation to such discrepancy may be due to the coupling between phonon excitation and the dynamic magnetic interaction, which follows the magnetic form factor. Hence, the dispersive excitation disappears at higher q value. 

We have fitted the data using random phase approximation (RPA) model.\cite{Broholm} Previously, the RPA model has been successful in describing inelastic phenomena in transition metal ion correlated systems.\cite{Birgeneau,Dahal} Fitting using RPA model is based on the assumptions that the only appreciable interaction is nearest neighbor interaction between Co-ions, $J$, and the interaction is isotropic in nature. The small ordered moment, despite the long counting time, and a nearly Q-independent excitation in MACS measurement conform to the applicability of RPA model to estimate the exchange interaction between magnetic ions. Under RPA model, $\chi$$^{''}$(Q,E) is described by,

\begin{eqnarray}
{\chi}^{^{\prime\prime}}(Q, \omega)&=&\sum_{\pm}\frac{\omega \chi_0 \Gamma_{Q^\pm}}{{\Gamma_{Q^\pm}}^{2}+\omega^2}
\end{eqnarray}

where $\Gamma$$_{Q^\pm}$ = $\Gamma$[1$\mp$$\chi_0$ $J$] and $\chi_0$ is static susceptibility. Clearly, the RPA model describes the dynamic properties of CoSeAs very well. Fitted value of $J$  = 0.25(4) meV at $T$ = 1.7 K indicates weak exchange interaction between Co-ions. Obtained values of $\Gamma$ is plotted as a function of wave vector Q in Fig. 4d. $\Gamma$, representing the full width at half maximum of dynamic correlation or the inverse of relaxation time $\tau$, seems to be independent of the wave vector Q. The dynamic susceptibility at $T$ = 135 K is barely distinguishable from the background. The small deviation from the background at low energy is most likely arising due to the paramagnetic fluctuation of Co ions at higher temperature.

\begin{figure}
\centering
\includegraphics[width=9.2cm]{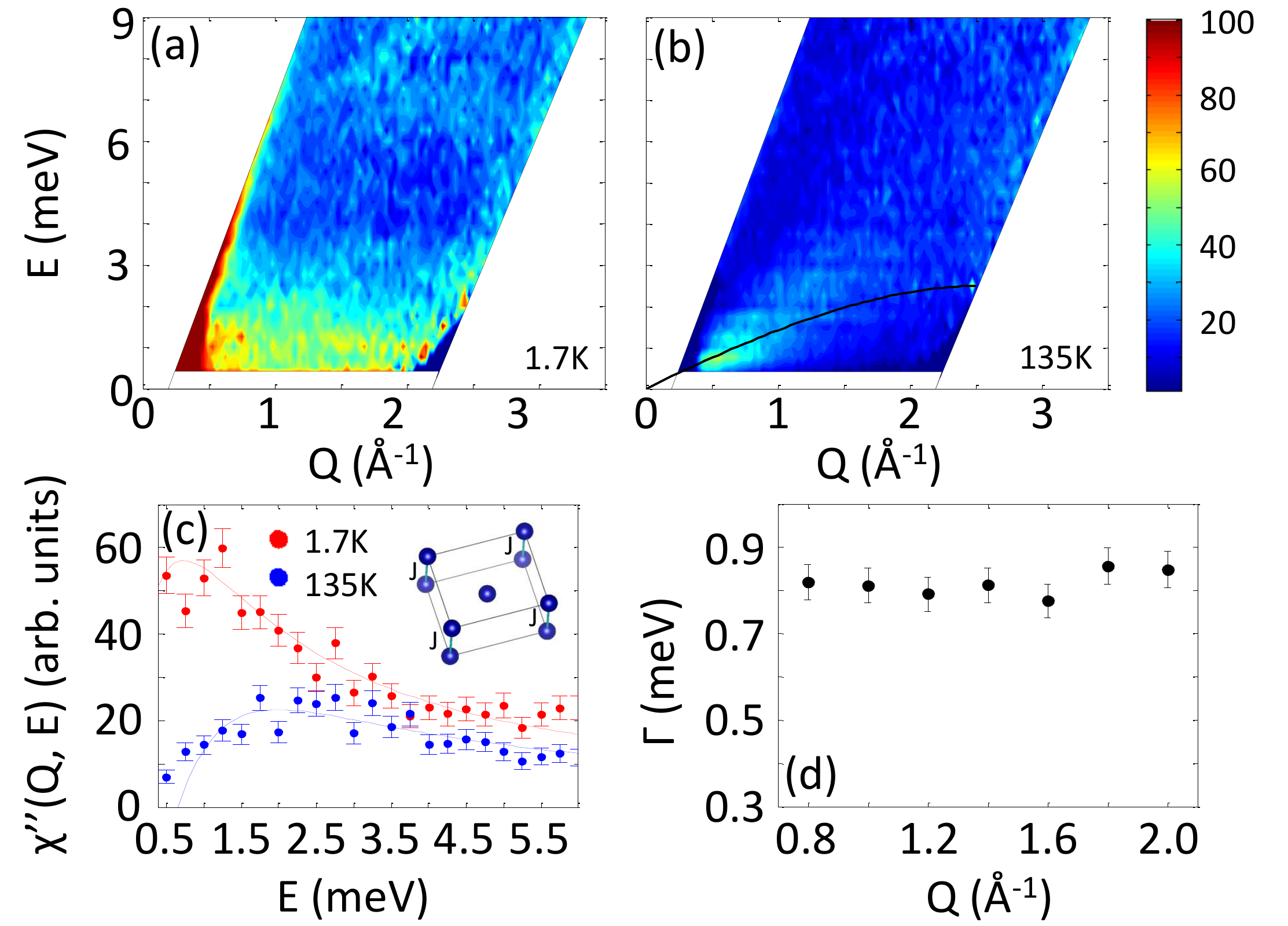} \vspace{-4mm}
\caption{(color online) Inelastic measurements of CoSeAs. (a-b) 2D map of energy-momentum, obtained on MACS spectrometer, at $T$ = 1.7 K and 135 K, respectively. Experimental data are background subtracted (by measuring Al can in identical experimental condition) and thermally balanced. Broad q-independent excitation is detected at $E$ $\leq$3 meV at low temperature. At a much higher temperature above magnetic ordering transition, the broad excitation disappears. Rather, a phonon dispersion tends to emerge (a curve is drawn for guide to the eye). (c) Plot of dynamic susceptibility vs. energy at fixed q = 1.4 $\AA$$^{-1}$ at low and high temperatures. Fitting using RPA model reveals clear peak-type structure at $E$ $\simeq$ 0.75 meV at $T$ = 1.7 K (see text for detail). (d) Estimated $\Gamma$ vs. q at $T$ = 1.7 K, manifesting q-independent characteristic of the dynamic behavior.
} \vspace{-4mm}
\end{figure}

The analysis of inelastic data reveals the large value of $\Gamma$, $\simeq$ 0.8 meV, for such a modest exchange coupled magnet. It suggests that Co ions are fluctuating with short relaxation time. Such behavior are usually observed in magnetically unstable systems.\cite{Ueland} Perhaps, the stoichiometric composition of CoSeAs is on the verge of a transition to another phase of material. 

In summary, we have synthesized a new marcasite phase material CoSeAs. Neutron scattering investigation of polycrystalline CoSeAs reveals the development of long range magnetic order below $T_c$ = 36.2 K. Given the fact that only a very few transition metal marcasites are known to manifest magnetic order, this is an important finding. Moreover, the weak nearest neighbor exchange coupling between Co ions makes it an interesting candidate material for the exploration of unconventional superconductivity using chemical doping method. Chemical pressure can further distort the anion octahedron, encompassing the transition metal ion, to induce a transition to new phase of material. Further research works are highly desirable in this pursuit. Besides the exploration of a possible superconducting state in chemically doped CoSeAs, we envision two possible applications of the new compound as photovoltaic absorber and in the design of robust electrocatalyst. There is an increasing trend in the use of marcasites and pyrites for photovoltaic application in recent years.\cite{Zhang,Wu} Future researches on the study of optical properties of CoSeAs thin film can elucidate its possible application in photovoltaics. More recently, an analogous marcasite CoSe$_{2}$ was demonstrated to preserve electrocatalytic integrity after long hours of usage in the acidic media.\cite{Zhang2} Similar studies on the crystalline specimen of CoSeAs can reveal new electrocatalytic properties in this compound.

DKS thankfully acknowledges the support by the Department of Energy, Office of Science, Office of Basic Energy Sciences under the grant no. DE-SC0014461. This work utilized facilities supported by the Department of Commerce.

\clearpage


\begin{thebibliography}{99}
\bibitem{NIST} NIST does not endorse any commercial product.
\end{thebibliography}
\end{document}